\documentclass{article}

\usepackage{arxiv}

\usepackage[utf8]{inputenc} 
\usepackage[T1]{fontenc}    
\usepackage{hyperref}       
\usepackage{url}            
\usepackage{booktabs}       
\usepackage{amsfonts}       
\usepackage{nicefrac}       
\usepackage{microtype}      
\usepackage{lipsum}		
\usepackage{graphicx}
\usepackage{amsmath}
\usepackage{doi}
\usepackage{comment}
\usepackage{caption}
\usepackage{cite}
\usepackage{subcaption}
\usepackage{authblk}
\usepackage[english]{babel}
\usepackage{csquotes}

\usepackage{subcaption}
\usepackage{xcolor}

\usepackage{pgfplots}
\usepackage{pgfplotstable}
\pgfplotsset{compat=1.17}
\usepackage{filecontents} 
\usepackage{pgfplots}
\usepgfplotslibrary{colorbrewer}
\pgfplotsset{compat = 1.15, cycle list/Set1-8} 
\usetikzlibrary{pgfplots.statistics, pgfplots.colorbrewer} 
\usepackage{pgfplotstable}
\usepackage{filecontents}

\newcommand{\bs}[1]{\boldsymbol{#1}}

\newcommand{\dpar}[2]{\frac{\partial #1}{\partial #2}}

\title{On the feasibility of foundational models for the simulation of physical phenomena}

\renewcommand{\shorttitle}{On the feasibility of foundational models for the simulation of physical phenomena}

\hypersetup{
pdftitle={On the feasibility of foundational models for the simulation of physical phenomena},
pdfsubject={q-bio.NC, q-bio.QM},
pdfauthor={E. Cueto},
pdfkeywords={First keyword, Second keyword, More},
}
\author[1]{Alicia Tierz$^{*,}$}
\author[1]{Mikel M. Iparraguirre$^{*,}$}
\author[1]{Iciar Alfaro}
\author[1]{David Gonz\'alez}
\author[2,3]{Francisco Chinesta}
\author[1]{El\'ias Cueto}
\affil[1]{{\small ESI Group-UZ Chair of the National Strategy on Artificial Intelligence. \protect\\ Aragon Institute of Engineering Research (I3A). Universidad de Zaragoza. Zaragoza, Spain.\protect\\ $^{*}$ Equal contributors.}}
\affil[2]{{\small CNRS@CREATE LTD. Singapore.}}
\affil[3]{{\small ESI Group chair. PIMM Lab. ENSAM Institute of Technology. Paris, France. }}

\begin{document}
\maketitle

\begin{abstract}
	We explore the feasibility of foundation models for the simulation of physical phenomena, with emphasis on continuum (solid and fluid) mechanics. Although so-called ``learned simulators'' have shown some success when applied to specific tasks, it remains to be studied to what extent they are able to undergo severe changes in domain shape, boundary conditions and/or constitutive laws and still provide robust (i.e., hallucination-free) and accurate results. In this paper we perform an exhaustive study of these features, put ourselves in the worst-case scenario and study their resistance to such strong changes in their domain of application.
\end{abstract}

\section{Introduction}

Learned simulators---those based on the use of machine learning tools---have attracted a great deal of interest in recent years because of their ability to simulate physical phenomena at a fraction of the cost of a traditional simulation \cite{stachenfeld2021learned,allen2022physical,yang2023learning,shlomi2020graph,heiden2021neuralsim,cicirello2024physics}. Of course, this cost must include the training process, but once the training process is developed, learned simulators are capable of delivering results under severe real-time constraints. This makes them suitable for repetitive tasks, for inverse problem solving or, in general, any task involving many queries.

In general, traditional simulators are based on the development of a mechanistic model (usually a partial differential equation or an ordinary differential equation) and its subsequent discretisation to give rise to an algebraic problem, solvable by a computer. Their success therefore depends on the quality of the starting model and its fit to the physics to be simulated, on the one hand, and on the computational cost resulting from the discretisation process, on the other.

Learned simulators, on the other hand, do not depend on the quality of the model, but on the quality of the data used for training. Their training cost is generally high, but once trained, during inference, they provide results very quickly.

The problem associated with such learned simulators is that, given the complexity usually associated with their loss function landscape, their predictions can fall into a local minimum and provide a physically meaningless result. Therefore, accuracy and robustness are a fundamental objective of the development of these methods.

To avoid these problems as much as possible, several authors have chosen to use so-called learning biases, methods that guide the prediction towards the global minimum of the cost function regardless of whether the input data are polluted by measurement noise. Of particular importance are the so-called inductive biases, those that are embedded in the architecture of the neural network \cite{cranmer2020discovering,battaglia2018relational}.

All these techniques have demonstrated high quality results, in terms of both accuracy and robustness. However, the big bottleneck they present right now is perhaps the lack of generality. These tools remain well suited to problems only slightly different from those for which they were trained. When the problem in question changes, either in terms of domain geometry, boundary conditions or material behaviour laws, the results provided still show important shortcomings that have prevented them, for the time being, from being the definitive alternative to traditional methods.

According to the Stanford Center for Human-centered Artificial Intelligence, ``Foundation models [...] are models trained on broad data at scale such that they can be adapted to a wide range of downstream tasks'' \cite{foundation}. Such models have proven to have phenomenal capabilities in the field of natural language processing \cite{zhou2023comprehensive} or medicine, to cite just a few of the possible applications \cite{huang2024foundation}. They are largely responsible for the popularity that artificial intelligence enjoys today. It would therefore seem logical to try to extend this philosophy to the field of simulation in the framework of continuum mechanics.

In this paper we have proposed a first analysis, to the best of our knowledge, of the possible application of the foundational model concept to the construction of general-purpose learned simulators. Aurora seems to be the only foundational model dedicated to simulation that has been developed so far. It is a weather simulation model. In this case, the adaptation of the model focuses not so much on changing the type of application for which it was developed (always weather forecasting) but on the particularisation of its predictions in the short or long term \cite{bodnar2024aurora}.

Caution should be exercised, first of all, in several respects. On the one hand, the foundational models used in natural language processing are trained on huge amounts of data. However, recent results seem to show that such a systematic search for bigger and bigger models is not always advisable \cite{varoquaux2024hypesustainabilitypricebiggerisbetter}. It seems that systematic growth in the size of models does not always correlate with a proportional improvement in results and that, at a certain size, bigger does not mean better. For many applications (e.g. medical image segmentation) it has been shown that models of an exaggerated size are not necessary. Moreover, as stated in \cite{besiroglu2024compute} ``a compute divide has coincided with a reduced representation of academic-only research teams in compute intensive research topics, especially foundation models''.

In the field of computational mechanics, access to such amounts of data is, for various reasons, not easy to obtain. On the one hand, the phenomena to be reproduced are complex, often involving strongly non-linear physics such as plasticity or contact. Simulating these problems often requires the design of costly simulation campaigns, which often show convergence failures, etc.

On the other hand, after the generation, curation, storage and analysis of such amounts of data are tasks that are often beyond the reach of academia, only accessible to large private companies, as has been the case with the aforementioned natural language processing models. Therefore, in general terms, the present work proposes a conservative line of work, in the sense that it will be adapted to the resources available in the context of a university with limited computing resources.

Our strategy will therefore be based on a worst-case scenario. Although the training of the architectures to be analysed hereafter has been made as general and extensive as possible, it should be noted that the results shown are based on a training that is in no way comparable to those currently in use in the field of natural language processing (NLP), for instance.

Therefore, by ‘worst case scenario’ we mean a strategy in which the training of our models is not as general as it could be, had we had the budget and computational means necessary for the realisation of one comparable to the NLP models.

Conversely, once the models to be analysed have been trained, the changes made in the inference phase will be as general as possible, but using the lowest possible computational cost. We have chosen to focus on the second part of the definition of a foundational model: ``they can be adapted to a wide range of downstream tasks''. For our purposes, these will include drastic changes in domain geometry, changes in boundary conditions and changes in constitutive laws. Models for both solid and fluid mechanics will be analysed.

The outline of the paper is as follows. Section \ref{methodology} reviews the general characteristics of the proposed models and the strategy followed to adapt them to changing end-use applications. The main emphasis is on the generality of the models, so that with only very slight refinements they can be adapted to purposes very different from those for which they were trained. Section \ref{examples} includes the above-mentioned examples of adaptation. Problems from solid (Section \ref{solid}) and fluid mechanics (Section \ref{fluid}) have been analysed.

In turn, within the examples dedicated to sound mechanics, two types of adaptations have been analysed. On the one hand, Section \ref{sec:mgn_plastic}, we have analysed the ability of the proposed architecture to withstand drastic changes, outside the distribution, both of the domain geometry and of the boundary conditions, for a problem involving contact and plasticity phenomena.

On the other hand, in Section \ref{sec:mgn_transfer} it has been analysed how this type of architecture can be modified if the needs change. Thus, for a system trained to provide the value of the von Mises stress, a transfer learning technique has been developed that allows, with very little additional training, the model to predict the complete stress tensor in the material.

The article is completed in Section \ref{conclusions} with a discussion of the possibilities offered by the strategies developed and an analysis of possible strategies for improvement.

\section{Methodology}\label{methodology}

Our approach will be based on the employ of graph neural networks \cite{Bronstein_2017,wu2020comprehensive,zhou2022graph}. Our interest comes from the fact that, on the one hand, the equations governing the mechanics of continuous media are local equations, and graph networks are able to reproduce this local structure very well and, on the other hand, we will use synthetic data from finite element and smooth particle hydrodynamics simulations, so that these data already incorporate a graph structure.

The use of graph networks to simulate the behaviour of physical systems is not new \cite{battaglia2018relational,sanchez2020learning,pfaff2020learning}. The fundamental advantage of graph networks is, on the one hand, the enormous speed of generating results in inference---one to two orders of magnitude faster than finite elements, for instance---and, on the other, the use they make of the mesh structure of the data, whether this is fixed, as in the case of solids, or adaptable, as in the case of the fluids shown here, in which a Lagrangian perspective has been chosen in the description of the physics. The particular architecture of each example is detailed in Section \ref{examples}.

Section \ref{solid}, which is devoted to solid mechanics, discusses examples that explore some of its most strongly non-linear phenomena: elasto-plasticity and contact.

In Section \ref{fluid}, which is devoted to free-surface fluid mechanics and the study of sloshing phenomena, it has also been decided to use inductive biases of the thermodynamic type \cite{Hernandez_2021,hernandez2022thermodynamics,moya2023thermodynamics,moya2024computational,urdeitx2024comparison}. These ensure compliance with the laws of thermodynamics (conservation of energy in closed systems and non-negative entropy production). As has been shown in the literature on this type of methods, they provide not only a higher accuracy in predictions, but also allow a smaller amount of data to be used than with traditional architectures, without thermodynamic biases.

In general terms, our system will be governed by a set of state variables that we group into a vector
$\bs z \in \mathbb R^{\tt n_{\text{part}}\times \tt n_{\text{sv}}}$, where $\tt n_{\text{part}}$ denotes the number of particles or nodes in the model and $\tt n_{\text{sv}}$ the number of state variables per node necessary to adequately describe the physics of the system. This, of course, depends on the particular example considered. Our simulators are designed to incrementally produce rollouts of the state of the systems for a number of discrete time increments,
$$
\bs Z =\{\bs z^0, \ldots, \bs z^t, \bs z^{t+\Delta t}, \ldots, \bs z^T  \}.
$$
A learned simulator actually computes the effect of the dynamics of the system over a time interval,
$$
\bs z^{t+\Delta t} = s_\theta (\bs z^t),
$$
where $s$ represents the simulator and $\theta$
 the set of learnable parameters of the model.

 All our architectures have a ``encode-process-decode'' structure. Once the mesh (or connectivity) structure of the model has been straightforwardly converted into a graph $G(V,E)$ composed by vertices $V$ and edges $E$ between them, there are a number of operations that are accomplished every time step. A first encoder computes nodal embeddings for each node, computing learned functions of their state. A different encoder computes pairwise embeddings for each pair of neighbouring particles or nodes. In the processing step, the system computes the interactions between vertices by means of a number of message passing steps, that transmits the information along the mesh. In our case, to observe Galilean invariance in the simulator, we store pairwise nodal relative distance at the edges. Finally, in the decoding step, information about the updated physical state of each vertex or edge is extracted. In the solid mechanics example, for instance, the von Mises equivalent stress is the variable extracted at each particle and time step.

In this work we have analysed different strategies for the adaptation of models to the simulation of previously unseen phenomena. 

In particular, we have studied the resilience to severe changes in domain geometry and boundary conditions (see Section \ref{sec:mgn_plastic}). The ability of the model to provide results under conditions very different from those under which it was developed is studied. To assess the generality of the trained model, we analyzed its zero-shot performance on out-of-distribution geometries and boundary conditions. 

In Section \ref{sec:mgn_transfer}, we study the application of transfer learning methods to the modification of model output variables. From a model trained to output von Mises stress, we use the trained backbone as a pre-trained foundation so as to provide the full second-order stress tensor at negligible computational cost. Only the decoder for the new output variables needs to be trained from scratch.

Finally, in the fluid sloshing example, Section \ref{fluid}, all training has been performed on the same container geometry (cylindrical). Adaptation is made to new container geometries at minimal computational cost (i.e., one-shot learning).

For each of the examples developed, details on the structure of the different networks that make up the model will be given in Section \ref{examples} below.

\section{Numerical examples}\label{examples}

The numerical examples are conducted within the computational mechanics domains of solid mechanics (Section \ref{solid}) and fluid mechanics (Section \ref{fluid}).  Despite the differences between these two domains---such as the physical phenomena, governing equations, and traditional computational mechanics solvers---the approach to solving them with machine learning is, at its core, the same: learning from data.

\subsection{Solid Mechanics: impact of an actuator on an elasto-plastic plate  }\label{solid}

As discussed in the introduction, the strategy of this paper is to take a worst-case scenario. Thus, we have chosen a problem involving contact and plasticity phenomena, two of the most strongly nonlinear phenomena we can find in solid mechanics. We consider a rectangular  plate with fixed geometric parameters across all sampled geometries of  length $L=0.5$ and width $W=0.25$. The young modulus $E=210$ and Poisson coefficient of $\nu = 0.3$. The yield stress is $300$ and the material is assumed to present hardening. The plate is impacted by a cylindrical actuator whose position along the length of the plate is a variable of the problem. The plate generally has about 600 nodes, although the numbers vary slightly for different findings depending on the position and size of the hole. The actuator has 450 nodes.
 
 The dataset consists of 135 high-fidelity three-dimensional finite element simulations, each containing 435 pseudo-time steps each, solved under quasi-static conditions. The collision is modeled as contact between an actuator (a rigid body) and a deformable plate (elastic-plastic material). All trajectories include loading and unloading, to capture the plasticity phenomena and ensure robustness under non-monotonic conditions. The variability in the dataset arises from changes in boundary conditions and geometry. The plates are perforated with holes of different sizes and positions. In addition, the thickness of the plate is also a variable.
 
 To this end, we extend MeshGraphNets (MGN) to simulate these phenomena, which---to the best of our knowledge---were previously limited to hyperelastic models \cite{pfaff2021learning}. This extension addresses higher nonlinearities and energy dissipation in the form of permanent deformations. 
 
The dataset is split into training, validation, and test sets as follows: 80 trajectories for training, 10 for validation, and 20 for testing. An additional test set, $\mathcal \mathcal \mathcal D_\text{extra}$ containing 25 trajectories with out-of-distribution geometries. 

\subsubsection{Modeling collisions in the plastic regime}
\label{sec:mgn_plastic}

\begin{figure}[h!]
    \centering
    \includegraphics[width=\textwidth]{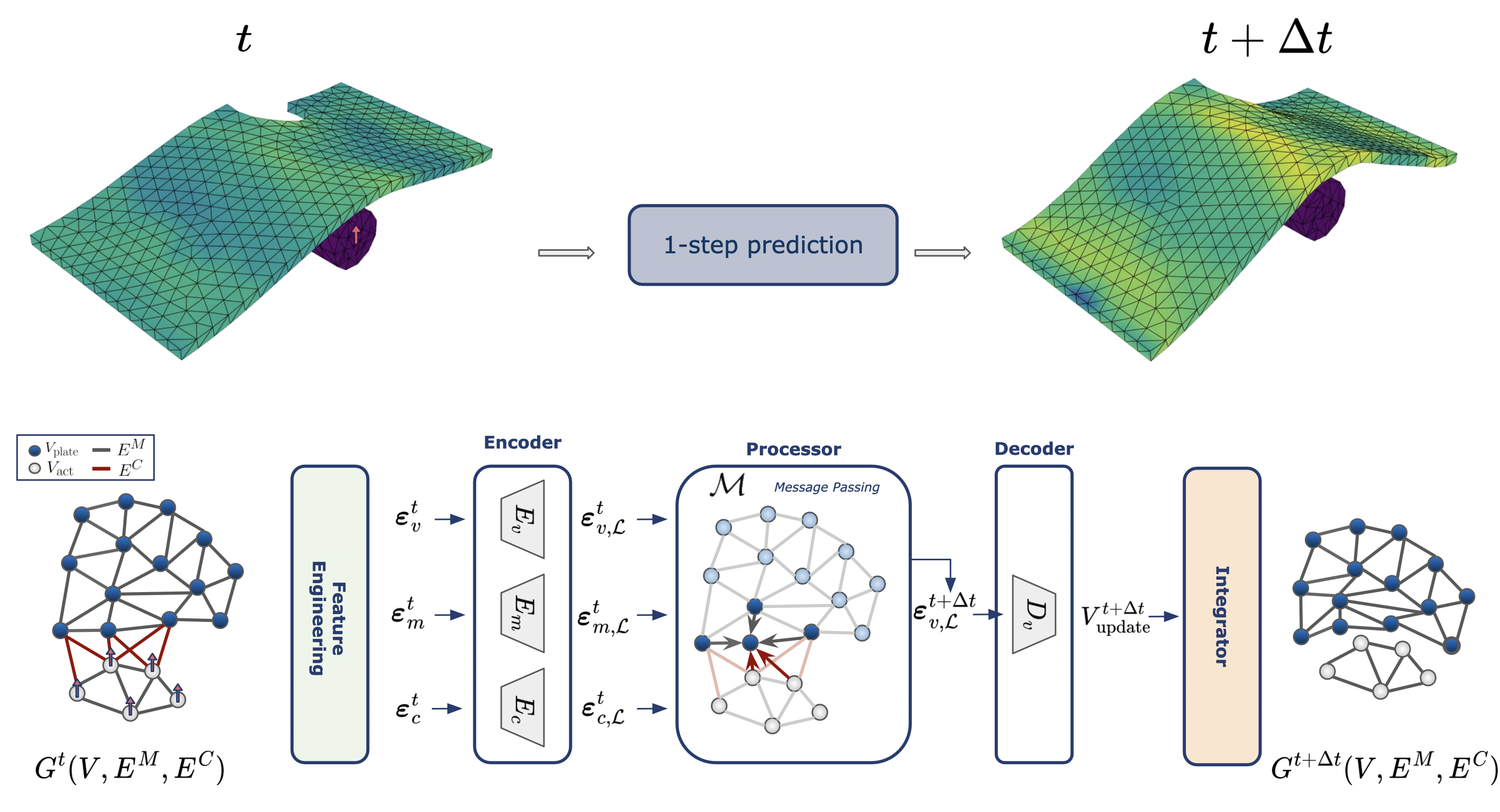}
    \caption{Pipeline illustration of the temporal integrator framework. Feature Engineering (green block), Encoder-Processor-Decoder (white blocks) and Integrator (orange block).}
    \label{fig:mgn_scheme}
\end{figure}

To model collisions in the plastic regime we trained the MeshGraphNet architecture from scratch in our dataset and the proposed temporal integrator framework \cite{pfaff2021learning}, where the model predicts the displacement and stress variables of each node for the next time step $t+\Delta t$, based on the positional information at $t$ and the imposed displacement by the actuator. MeshGraphNets follow the encoder-processor-decoder structure sketched in Fig. \ref{fig:mgn_scheme} where, despite of being a node-to-node architecture,  it can update via message passing the graph to perform global predictions for all the nodes.

To model the physical system at each time step $t$ we establish a one-to-one correspondence between the finite element mesh and an undirected graph $G^t = (V, E^M)$, typical of graph neural networks, where $V$ represents the set of $\tt n_{\text{part}}$ vertices containing the state variables $\bs z$ and $E^M$ are the edge connections of the mesh nodes  from the simulation, of both plate and actuator $G_{\text{plate}}^t = (V_{\text{plate}}, E_{\text{plate}}^M),  \;\;G_{\text{act}}^t = (V_{\text{act}}, E_{\text{act}}^M)$. 

In order to model the plate-actuator contact, an extra set of edges must be created based on a radial contact distance $r_C$, named as contact edges $E^C$. If the euclidean distance $|\bs q_i - \bs q_j| < r_C$, then nodes $i$ and $j$ will be connected. These edges only connect the actuator to the plate and are uni-directed, to guarantee the information flow from actuator to plate. Now the system is represented as a multi graph $G(V, E^M, E^C)$.

\paragraph{Feature Engineering:} All the positional information is encoded as edge attributes as a geometric inductive bias to make the model translation-invariant. The mesh edge attributes $\bs \varepsilon^t_{m, ij}$ are the concatenation of the distance and its norm between mesh nodes $\bs{d}^t_{ij} = ( \bs{q}^t_i - \bs{q}^t_j, |\bs{q}^t_i - \bs{q}^t_j|)$ for the time $t=0$ (undefomed) and the given time $t$ (deformed), resulting in $\bs \varepsilon^{t}_{m, ij} = (\bs{d}^{0}_{ij}, \bs{d}^{t}_{ij}) \quad \forall \, i,j \in {E}^M$. The incorporation of  both distances together helps the network to learn the sense of physical deformation across edges. The contact edges attributes are $\bs \varepsilon^{t}_{c, ij} = (\bs{d}^{t}_{ij}) \quad \forall \, i,j \in {E}^C$, where only the distance between actuator and mesh nodes is required.
The node attributes are the concatenation of the node type $\bs{n}_i$ as one-hot-vector and the imposed displacement for the actuator nodes $\bs{u}_{\text{actuator}}^{t} = (\bs{q}^{t+\Delta t} - \bs{q}^t$), being always set to zero for the plate nodes $\bs{u}_{\text{plate}}^{t} = \bs{0}$, resulting in $\bs \varepsilon^{t}_{v,i} = (\bs{n}_i, \bs{u}_i^{t}) \quad \forall \, i,j \in V $. See details in Fig. \ref{fig:mgn_scheme}.

\paragraph{Encoder-Processor-Decoder architecture:}
\begin{itemize}
    \item \textbf{Encoder}: its role is to project the features to the high-dimensional latent space $\mathcal{L}$. Three different 2-layer MLPs with 128 hidden units and LeakyReLU activation function are used as encoders. The node encoder $E_v$ for the nodes attributes $\bs \varepsilon^{t}_{v,i}$, the edge mesh encoder $E_m$ for the edge mesh attributes $\bs \varepsilon^{t}_{m, ij}$ and the edge contact encoder $E_c$ for the edge contact attributes.

    \item \textbf{Processor}: is in charge of updating all the latent variables, node, edge mesh and edge contact attributes via message passing. This part of the architecture learns the actuator-plate interaction and the response of the plate to such interaction dependent on the material's behavior implicitly learnt form the data. The message passing consists on $M=15$ message passing blocks of 2-layer MLPs with 128 hidden units and LeakyReLU activation.
    
    \item \textbf{Decoder}: decodes the update variables $V^{t+\Delta t}_{\text{update}}$ from the node attributes $\bs \varepsilon^{t+\Delta t}_{v,\mathcal{L},i}$ as output of the processor block. In this section the output variables are displacements and von Mises stress $(\bs{u}^{t+\Delta t}_i, \sigma^{t+\Delta t}_{\text{VM}, i})$ for each node of the plate, $i \in V_{\text{plate}}$.

    \item \textbf{Loss function}: The computed loss is the mean squared error between the target variables $(\hat{\bs{u}}_i,  \hat{\sigma}_{\text{VM}, i})$ and the ones predicted by the model  $(\bs{u}_i,\sigma_{\text{VM}, i})$, 
    \begin{equation}
    \text{Loss} = \frac{1}{\tt n_{\text{plate}}} \sum_{i=1}^{\tt n_{\text{plate}}} \left( (\bs{u}_i - \hat{\bs{u}}_i)^2  + ( {\sigma}_{\text{VM}, i} - \hat{\sigma}_{\text{VM}, i})^2\right).
    \label{eq:loss}
    \end{equation}
    \end{itemize}

\textbf{Integrator:}  
This part enables rollouts using a one-step prediction model by integrating the output $ V^{t+\Delta t}_{\text{update}}$  into $ G^{t}(V, E^M, E^C) $ to obtain $ G^{t+\Delta t}(V, E^M, E^C) $. The predicted displacements are integrated to update the positions of the graph nodes as \( \bs{q}^{t+\Delta t} = \bs{q}^{t} + \bs{u}^{t+\Delta t}_i \), while the von Mises stress variable \( \sigma^{t+\Delta t}_{\text{VM}, i} \) is directly decoded. Normal nodes are updated according to this scheme, while fixed and actuator nodes are constrained to their respective boundary condition values.

\textbf{Experiments}

 To assess the model's performance, we measure the mean error during rollouts at $n=50$ steps and $n=450$ steps for both the positions $\bs q$ and von Mises stress $\sigma_{\text{VM}}$ variables for all nodes $V$. The metrics used are the Root Mean Squared Error, 
 \begin{equation}
\text{RMSE} = \sqrt{\frac{1}{n}\sum_{i=1}^{n} \| \bs y_{i} - \hat{\bs y}_{i} \|^2},
\label{eq:rmse}
\end{equation}
 and the Relative Root Mean Squared Error, as given by 
 \begin{equation}
\text{RRMSE} = \sqrt{\frac{1}{n} \sum_{i=1}^{n} \left( \frac{\| \bs y_i - \hat{\bs y}_i \|}{\| \hat{\bs y}_i \|_{\infty}} \right)^2},
\label{eq:rrmse_inf}
\end{equation}
where $\bs y$ represents the output variable, either $\bs q$ or $\sigma_{\text{VM}}$.
 The error bars denote the standard error across different trajectories.

 The first part of the research focuses on evaluating the model's zero-shot performance, assessing its ability to handle unseen boundary conditions and geometries without any fine-tuning. This evaluation is crucial for determining the range of performance and the limitations of the trained models.

To enhance the model's versatility, we designed a diverse set of trajectories that explore different boundary conditions and geometries. Each trajectory is distinct, sampled from the specified distribution of parameters:
\begin{itemize}
    \item Boundary conditions: Either left, right of both sides of the plate can be clamped. The actuator is located at a position \(\bs x \in [0.25 L_{\text{plate}}, 0.75 L_{\text{plate}} ]\) across plate length and the imposed displacement of the actuator \( u_{\text{act}} \in [-0.1, 0.1]\).
    \item Plate geometry: The same applies to the geometry parameters. The thickness of the plate, \(T_{\text{plate}} \in [0.03, 0.1]\). Hole radius \(H^{\text{r}}_{\text{plate}} \in [0.05, 0.1]\) and hole location \(H^{\text{loc}}_{\text{plate}} \in \mathcal{D}_{\text{plate}}\).
\end{itemize}


While $\mathcal{D}_{\text{train}}$, $\mathcal{D}_{\text{valid}}$ and $\mathcal{D}_{\text{test}}$ are trajectories designed by sampling from these distributions, the ones of $\mathcal{D}_{\text{extra}}$ sample the hole radius in the out-of-distribution interval   \( [0.1, 0.15]\).

\begin{figure}[h!]
    \centering
    \includegraphics[width=\textwidth]{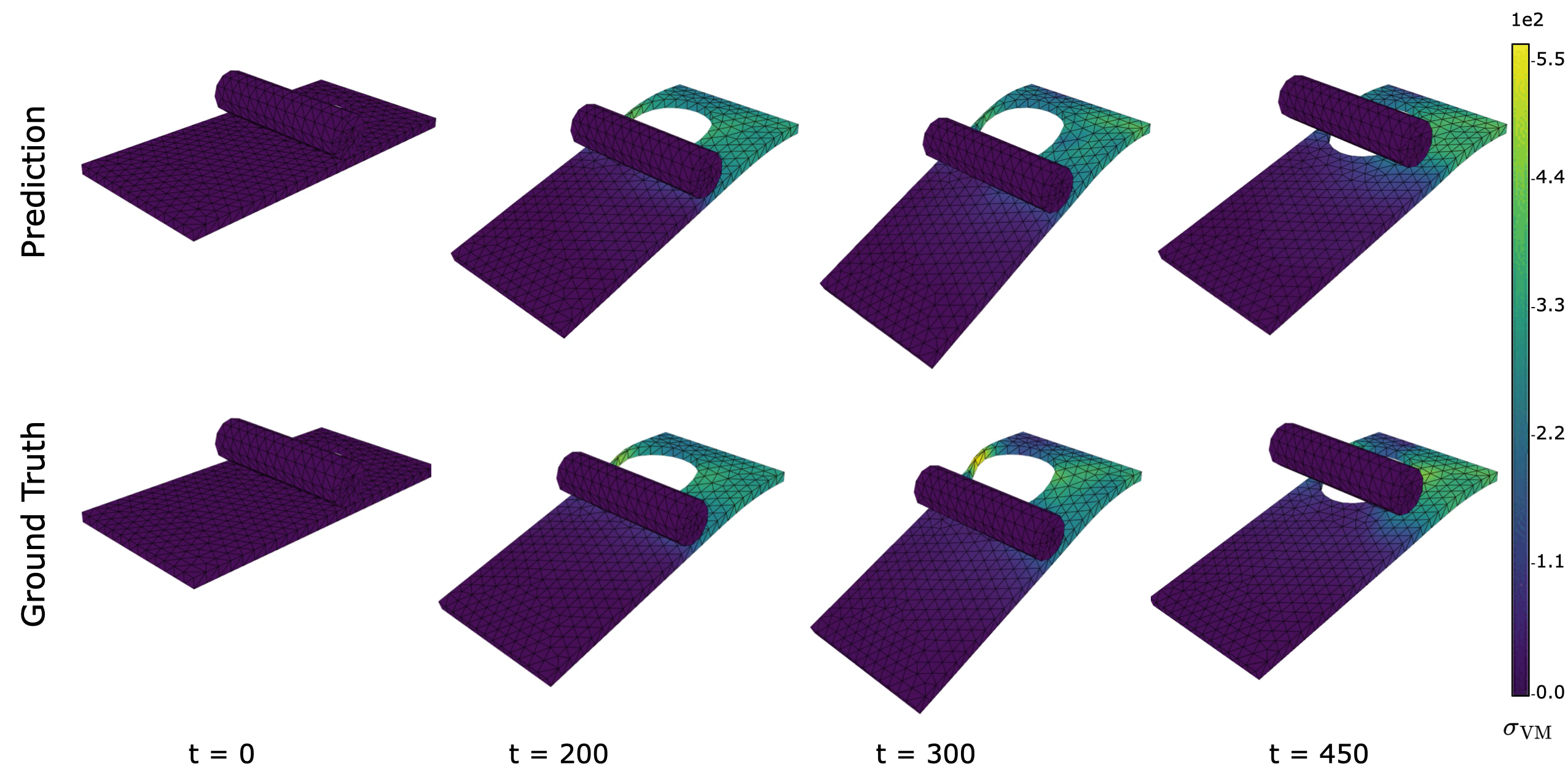}
    \caption{Snapshots of the rollout for trajectory in $ \mathcal D_{\text{test}}$, unseen geometry and boundary conditions. Loading was applied from $t=0$ to $t=330$, while unloading occurred from $t=330$ to $t=450$. {Top}: prediction. {Bottom}: ground truth. }
    \label{fig:test_vmises_rollout}
\end{figure}

\begin{figure}[h!]
    \centering
    \includegraphics[width=\textwidth]{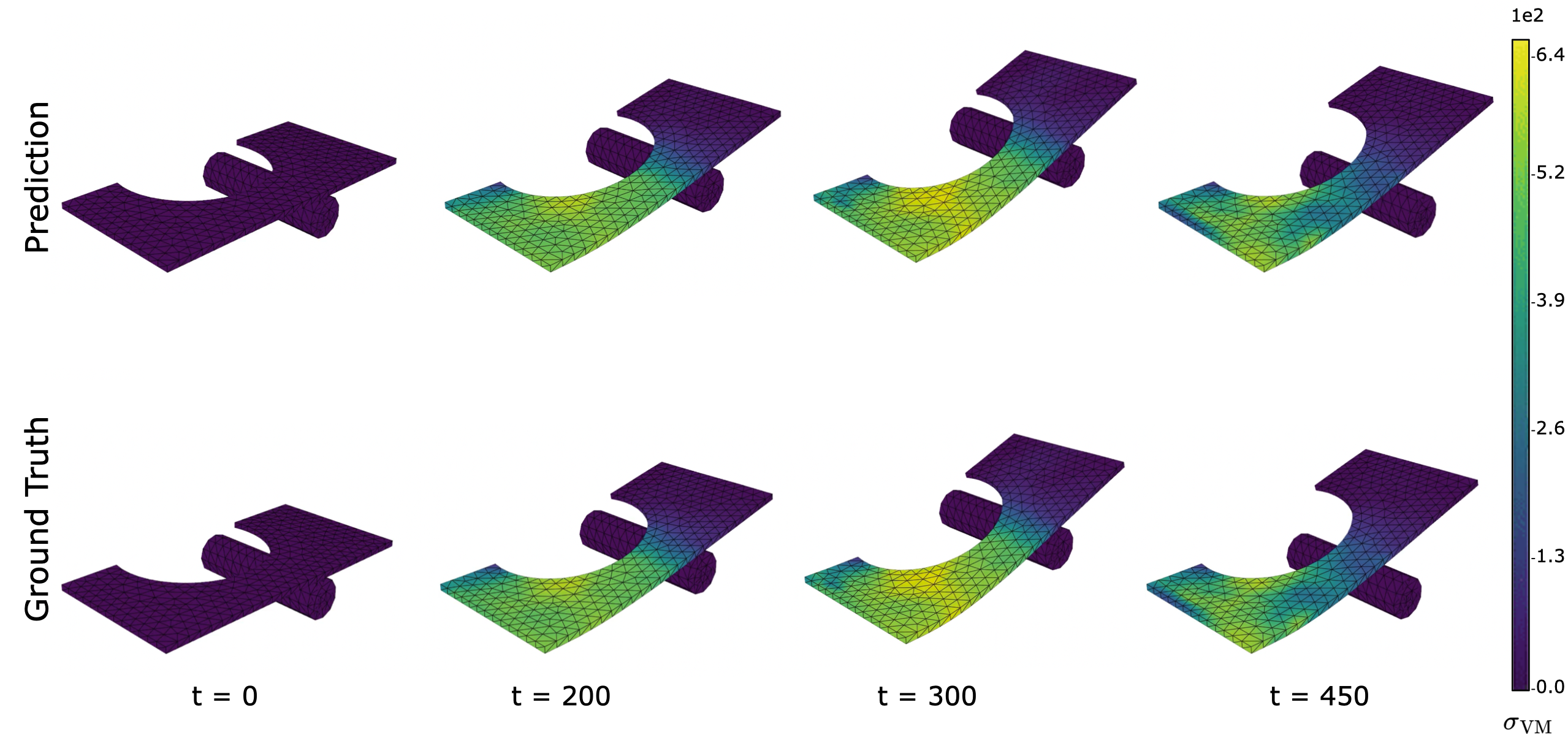}
    \caption{ Snapshots of the rollout for trajectory in $ \mathcal D_{\text{extra}}$, unseen boundary conditions and geometry out-of-distribution. Loading was applied from $t=0$ to $t=330$, while unloading occurred from $t=330$ to $t=450$. {Top}: prediction. {Bottom}: ground truth.}
    \label{fig:test_extra_vmises_rollout1}
\end{figure}

The behaviour of the model for the different test splits is summarised on Table \ref{tab:rmse_rrmse}. Visual results are presented in Figs. \ref{fig:test_vmises_rollout} and \ref{fig:test_extra_vmises_rollout1}.

\begin{table}[h]
    \centering
    \begin{tabular}{|c|c|c|c|c|}
        \hline
        \textbf{Variable} & \textbf{50-step, $\mathcal{D}_{\text{test}}$} & \textbf{50-step, $\mathcal{D}_{\text{extra}}$} & \textbf{450-step, $\mathcal{D}_{\text{test}}$} & \textbf{450-step, $\mathcal{D}_{\text{extra}}$} \\
        \hline
        {RMSE (\( \bs q \)) \(\times 10^{-3}\)}  & 1.715 $\pm$ 0.364 & 2.193 $\pm$ 0.377 & 6.579 $\pm$ 1.290 & 15.352 $\pm$ 2.720 \\
        \hline
        {RMSE (\( \sigma_\text{VM}\)) \(\times 10^{3}\)} & 34.047 $\pm$ 6.020 & 44.048 $\pm$ 5.349 & 50.924 $\pm$ 9.078 & 63.675 $\pm$ 6.873 \\
        \hline
        {RRMSE (\( \bs q \))(\%)} & 0.3862 $\pm$ 0.08212 & 0.4936 $\pm$ 0.0848 & 1.350 $\pm$ 0.2634 & 3.0956 $\pm$ 0.541  \\
        \hline
        {RRMSE (\( \sigma_\text{VM}\))(\%)} & 33.510 $\pm$ 6.970 & 21.263 $\pm$ 2.302 & 15.001 $\pm$ 2.845 & 15.055 $\pm$ 1.461 \\
        \hline 
    \end{tabular}
    \vspace{0.3cm}
   \caption{Zero-shot performance under unseen boundary conditions and geometries. RMSE and RRMSE values for position (\(q\)) and von Mises stress (\(\sigma_\text{VM}\)) at different rollout steps (50 and 450, respectively) for $\mathcal{D}_{\text{test}}$ and $\mathcal{D}_{\text{extra}}$ datasets.  The mean was computed across all the nodes, time steps and trajectories.}
    \label{tab:rmse_rrmse}
\end{table}


\begin{figure}[h!]
    \centering
    \includegraphics[width=\textwidth]{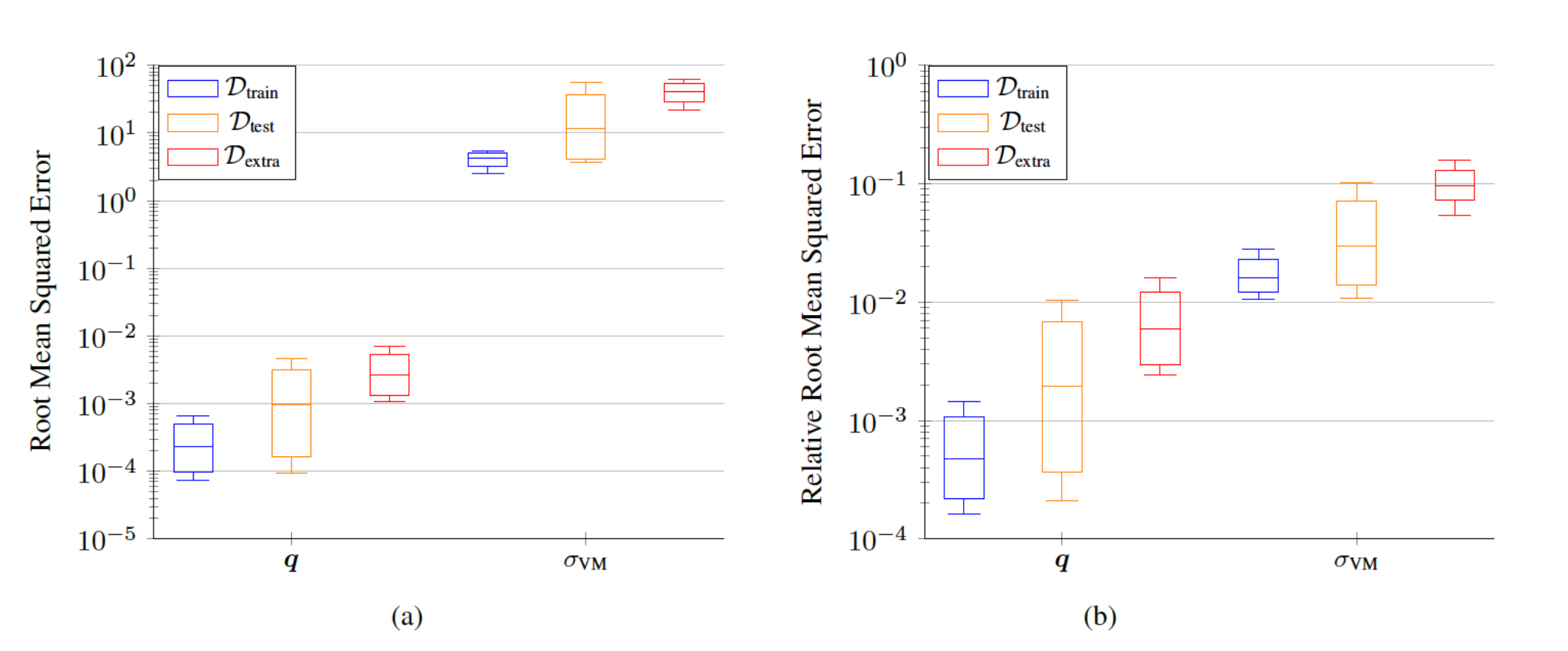}
        \caption{The box plots display the distribution of the accumulated error per trajectory over 450-step rollouts. Each point on the plot represents the error an individual trajectory. a) Root mean squared error for position $\bs{q}$ and von Mises stress $\sigma_{\text{VM}}$.  b) Relative root mean squared error for position $\bs{q}$ and von Mises stress $\sigma_{\text{VM}}$. }
\end{figure}

This method is quite data-hungry and requires long training times. The model was trained on a single RTX 4090 GPU for 1k epochs with a batch size of 4, where each epoch takes around 10k steps to be completed. The training process has a wall-clock time cost of approximately two weeks. It is crucial the impact of adding noise, which serves as an augmentation technique in the training process to enhance robustness during rollout, as the target value for displacements $\bs{u}$ will be computed as $\hat{\bs{u}}^{t+\Delta t} = \bs{q}^{t+\Delta t} - \bs{q}^{t}_{\text{noisy}} $.

\subsubsection{Changing output variables through transfer learning}
\label{sec:mgn_transfer}

\begin{figure}[h!]
    \centering
    \includegraphics[width=\textwidth]{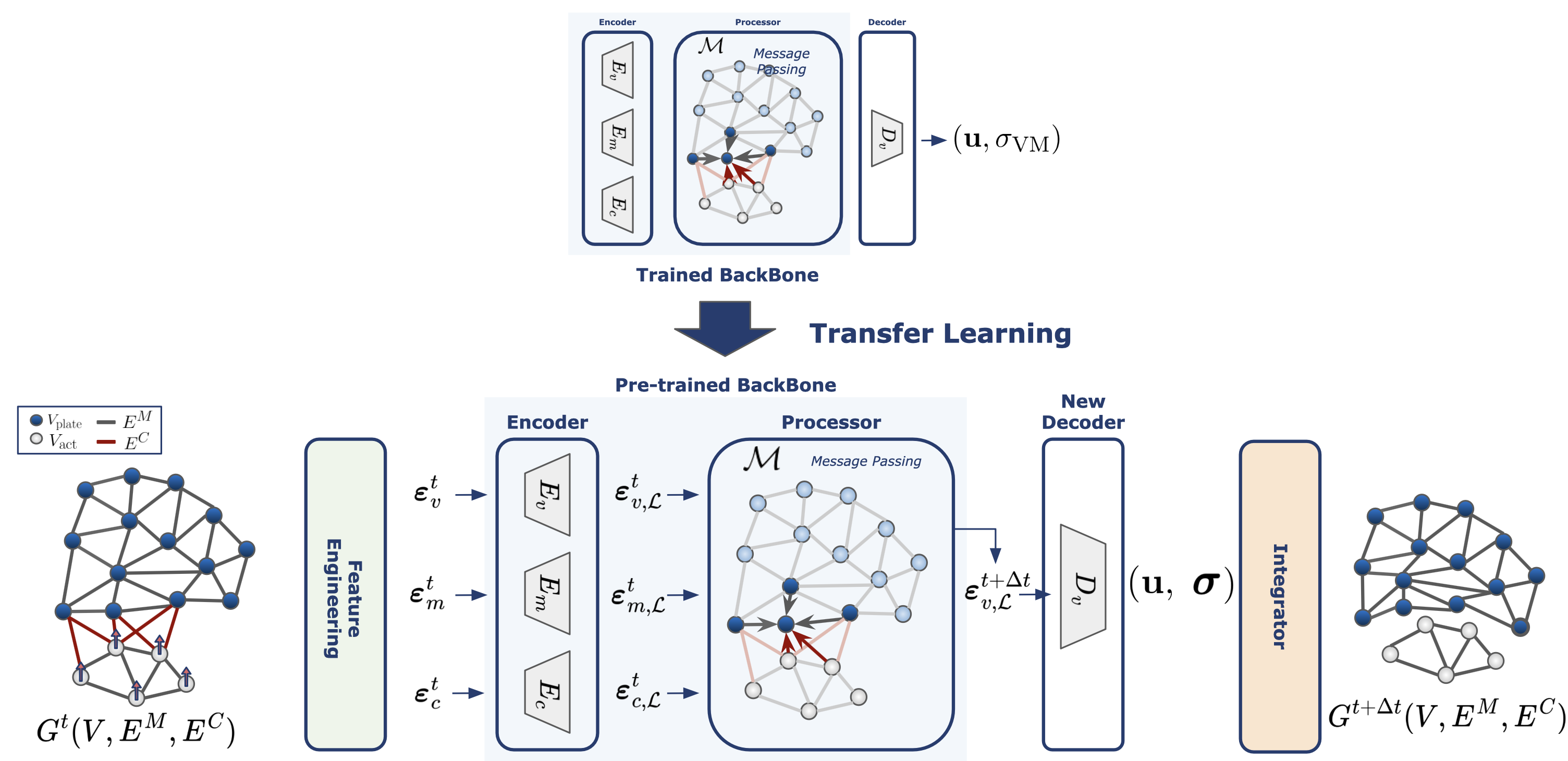}
    \caption{Sketch of the transfer learning approach. A model trained to provide von Mises stresses as output variable is modified so as to provide the whole stress tensor field at a negligible cost. }
    \label{fig:transferlearning}
\end{figure}

In this section we study how to modify the model in the hypothetical situation that we need to require new output variables, different from those for which it was trained.

Up to this point, the Encoder-Processor has learned the actuator-plate interaction and the behavior of the plates in the latent space, from which the decoder simply decodes the displacements \(\bs{u}\) and \(\sigma_\text{VM}\), which were the variables initially required of the model. Therefore, we hypothesize that these latent node variables should also embed valuable information about the complete second-order stress tensor \(\bs\sigma\) due to their correlation with the \(\sigma_\text{VM}\) variable. This presents an alternative to training the models from scratch, which does take advantage of the already learned patterns, such as the actuator-plate interactions.

\textbf{Experiments}:

 We transferred the learned encoder-processor weights to a new architecture, where the decoder decodes from the latent space $\mathcal{L}$, $V^{t+\Delta t}_{\text{update}}$  to be the displacements $\bs u$ and the six components of the stress tensor $\bs{\sigma}$, see Fig.  \ref{fig:transferlearning}. In order to improve convergence, the training process is handled in two stages: 
 \begin{enumerate}
     \item Train the model with freezed  encoder-processor weights only updating the new decoder and
     \item  train the whole model with a learning rate of $ 10^{-5}$.
 \end{enumerate} 

This resulted in considerable speed-ups ($\times 10$) in training time, achieving convergence after only 50 epochs. These results are summarized in Table \ref{tab:rmse_rrmse_transfer}.

\begin{table}[h]
    \centering
    \begin{tabular}{|c|c|c|c|c|}
        \hline
        \textbf{Variable} & \textbf{50-step, $\mathcal{D}_{\text{test}}$} & \textbf{50-step, $\mathcal{D}_{\text{extra}}$} & \textbf{450-step, $\mathcal{D}_{\text{test}}$} & \textbf{450-step, $\mathcal{D}_{\text{extra}}$} \\
        \hline
        {RMSE (\( \bs q \)) \(\times 10^{-3}\)}  & 1.742 $\pm$ 0.327 & 2.285 $\pm$ 0.349 & 7.824 $\pm$ 1.287 & 16.372 $\pm$ 2.70 \\
        \hline
        {RMSE ($\bs{\sigma}$) \(\times 10^{3}\)} & 40.747 $\pm$ 2.857 & 39.871 $\pm$ 2.909 & 91.099 $\pm$ 6.506 & 89.541 $\pm$ 6.913 \\
        \hline
        {RRMSE (\(\bs q \))(\%)} & 0.391 $\pm$ 0.0736 & 0.511 $\pm$ 0.07787 & 1.611 $\pm$ 0.25993 & 3.317 $\pm$ 0.5357 \\
        \hline
        {RRMSE($\bs{\sigma}$)(\%)} & 27.312 $\pm$ 4.631 & 24.228 $\pm$ 3.117 & 10.169 $\pm$ 1.447 & 9.840 $\pm$ 0.991 \\
        \hline
    \end{tabular}
    \vspace{0.3cm}
   \caption{Zero-shot performance under unseen boundary conditions and geometries after transfer learning. RMSE and RRMSE values for position (\(\bs q\)) and second-order stress field $\bs{\sigma}$ at different rollout steps (50 and 450, respectively) for $\mathcal{D}_{\text{test}}$ and  $\mathcal{D}_{\text{extra}}$ datasets. }
    \label{tab:rmse_rrmse_transfer}
\end{table}

\begin{figure}[h!]
    \centering
    \includegraphics[width=\textwidth]{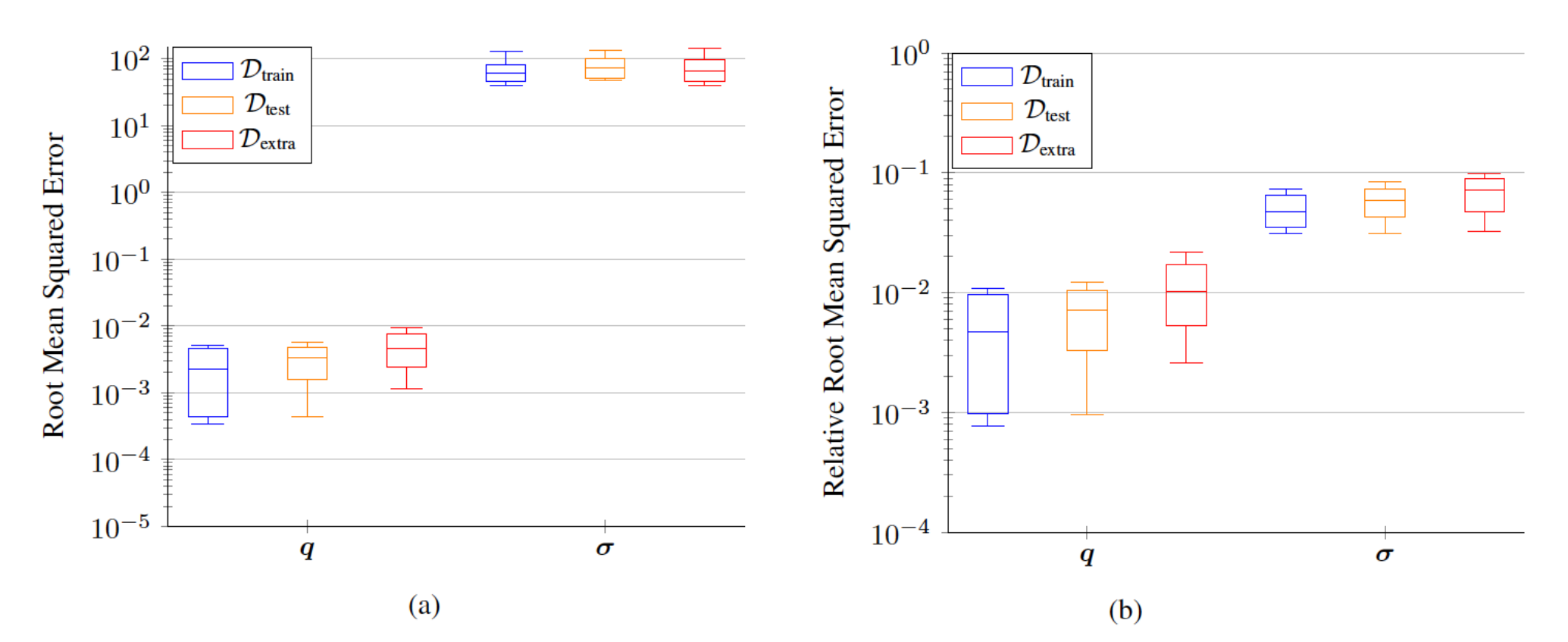}
    \caption{The box plots display the distribution of the accumulated error per trajectory over 450-step rollouts. Each point on the plot represents the error an individual trajectory. a) Root mean squared error for position $\bs{q}$ and stress field $\bs \sigma$.  b) Relative root mean squared error for position $\bs{q}$ and von Mises stress $\sigma_{\text{VM}}$. }
\end{figure}

\begin{figure}[h!]
    \centering
    \includegraphics[width=\textwidth]{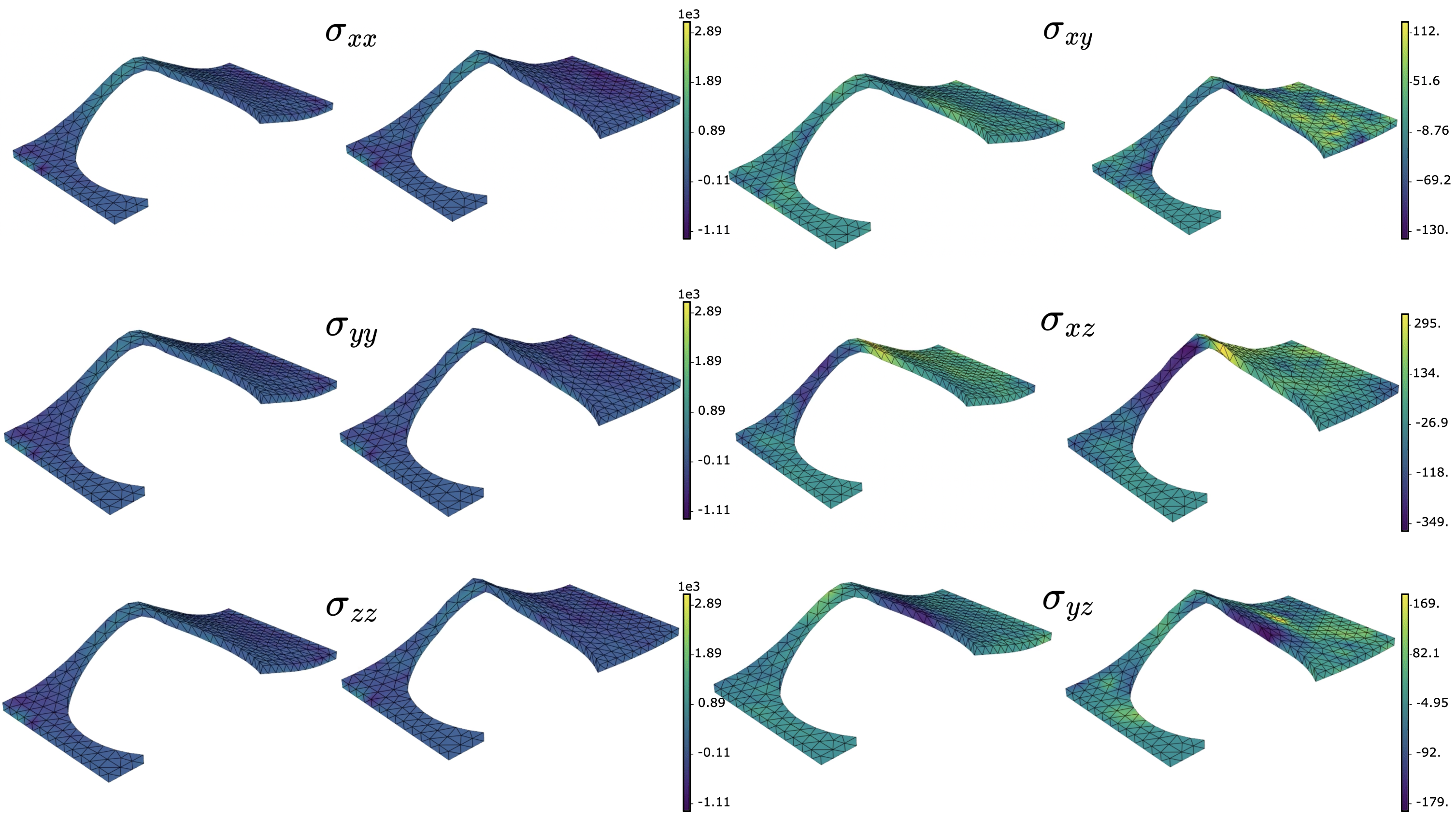}
    \caption{Example of the second-order stress variables $\bs{\sigma}$ predictions for a rollout at step 450 for trajectory $\in \mathcal D_{\text{extra}}$. Every pair of figures represent a component variable.  For each one, the left column includes predicted values, while the right ones represent the ground truth. For the sake of clarity, only the stress variables are plotted, while the actuator is omitted from the visualization}
    \label{fig:test_extra_vmises_rollout}
\end{figure}

\subsection{Fluid Mechanics}\label{fluid}

In the examples above, a Lagrangian description, typical of solid mechanics, and a description based on the use of a fixed mesh has been used (the possibility of remeshing has not been considered). In this section, on the contrary, an equally (updated) Lagrangian description is employed, which involves the need to use a continuous process of remeshing, i.e. of recomposing the connectivity of each particular material describing the fluid. This will make it possible to demonstrate the generality of the proposed methodology.

In this third numerical example, we propose to leverage foundational models and transfer learning to address challenges related to geometry change in the fluid container. A model has been trained on a specific geometry (in this case, a cylindrical container with different filling levels) and, by applying transfer learning, the model has quickly adapted to new geometries with very short training times. The central idea is that the underlying physical principles, such as the Navier-Stokes equations, remain constant in all geometries, allowing the model to learn these fundamental dynamics and apply them to new configurations. Again, this idea is related to the local character of the architecture of graph networks, with encoders and decoders shared by each vertex and by each edge.

\subsubsection{Geometric and metriplectic structure}

To address the challenge of efficiently training models across diverse container geometries, we employ a thermodynamics-informed graph neural network (TIGNN) architecture that incorporates both geometric and inductive biases to enhance its generalisation capabilities \cite{hernandez2022thermodynamics}. The geometric bias of TIGNN derives from its inherent invariance to node rotations and translations, ensuring that model performance remains robust regardless of the spatial orientation of the input geometries. This property is crucial for fluid simulations, where the geometry of the domain can be varied without changing the underlying physical laws.
The structure of the graph $G^t = (V, E^M)$, is similar to the two previous numerical examples, but with one key difference: there is no actuator present. The graph contains two types of nodes: those belonging to the container, which are fixed and cannot be predicted, and those belonging to the fluid, which are mobile. There is only one type of edge in the graph, which connects nodes that are within a given radius and its properties include the relative distance $(\bs q_{ij} = \bs q_{i} - \bs q_{j})$ between the connected nodes.

In addition to the geometric bias, we introduce an inductive bias into the loss function by applying the GENERIC (General Equation for the Non-Equilibrium Reversible-Irreversible Coupling) formalism \cite{GENERIC}. This mathematical framework is specifically designed to encode the underlying thermodynamic structure of the problem, providing the model with additional information about the physics governing fluid dynamics. This additional information, in the absence of any other detail about the precise form of the problem, refers to the fulfillment of the principles of thermodynamics. By assuming that the fluid will flow according to the dictate of GENERIC (a completely general metriplectic formalism), we guide the model towards solutions that respect the fundamental conservation laws of fluid mechanics, thus improving both the accuracy and physical consistency of the predictions.

Considering a fluid mass discretized into $\tt n_v$ particles. To enforce the fulfillment of the laws of thermodynamics, we assume a metriplectic evolution in the fluid. The objective of the training phase will be to unveil its precise form:
\begin{equation}\label{GENERIC2}
\dot{ \bs z} = \bs L( \bs z) \frac{\partial E}{\partial \bs z} + \bs M( \bs z)\frac{\partial S}{\partial \bs z},
\end{equation}
where $\bs z \in \mathbb{R}^{\tt n_v \times \tt n_{\text{dof}}}$ and $\tt n_{\text{dof}}$ represents the number of degrees of freedom of each particle. $\bs L$ represents a simplectic (Poisson) matrix, which is known to be skew-symmetric, and $\bs M$ is a symmetric and positive semi-definite dissipation matrix. $E$ and $S$ represent the energy of the system and a generalized entropy potential.

The fulfillment of the first and second principles of thermodynamics, given this metriplectic structure for the problem, is ensured  if the so-called degeneracy conditions are satisfied, i.e.,
\begin{equation}\label{deg1}
\bs L(\bs z)\frac{\partial S}{\partial \bs z} = \bs 0,
\end{equation}
and
\begin{equation}\label{deg2}
\bs M(\bs z)\frac{\partial E}{\partial \bs z} = \bs 0.
\end{equation}
This holds since
\begin{equation}\label{eq:Econs}
 \dpar{E}{t} = \dpar{E}{\bs{z}}\cdot\dpar{\bs{z}}{t}=\dpar{E}{\bs{z}}\left( \bs{L} \dpar{E}{\bs{z}} + \bs{M} \dpar{S}{\bs{z}}\right)=0,
\end{equation}
which is the proof of the conservation of energy in an isolated system---in other words, the first law of thermodynamics---. A similar reasoning is established for the entropy $S$, which satisfies:
\begin{equation}\label{eq:Scons}
 \dpar{S}{t} = \dpar{S}{\bs{z}}\cdot\dpar{\bs{z}}{t}=\dpar{S}{\bs{z}}\left( \bs{L} \dpar{E}{\bs{z}} + \bs{M} \dpar{S}{\bs{z}}\right)=\dpar{S}{\bs{z}}\bs{M}\dpar{S}{\bs{z}}\geq 0,
\end{equation}
which is the sought enforcement of the second law of thermodynamics.

In order to avoid the assembly of big matrices, the GENERIC formalism we employ is a local implementation of thermodynamics-informed GNNs \cite{tierz2024}. Instead of enforcing the principles of thermodynamics for the whole system, considered as a closed one, we assume that each fluid particle is an open thermodynamic system. Consider a particle $i$ with state variables $\bs z_i \in \mathbb{R}^{\tt n_{\text{dof}}}$, connected in the graph to $j=1, \ldots, \tt n_{\text{neigh}}$ neighboring particles. The state variables of particle $i$ evolve over time according to the following equation:
\begin{equation}
\dot{\bs z}_i = \bs L_i (\bs z_i)  \frac{\partial e_i}{\partial \bs z_i} + \bs M_i(\bs z_i)\frac{\partial s_i}{\partial \bs z_i}  
- \sum_j^{\tt n_{\text{neigh}}} \left[ \bs L_{ij} (\bs z_j)\frac{\partial e_j}{\partial \bs z_j} + \bs M_{ij}(\bs z_j) \frac{\partial s_j}{\partial \bs z_j} \right], 
\label{eq:port2}
\end{equation}
where the evolution is influenced by both local and neighboring particles through the terms involving $\bs L$ and $\bs M$ matrices.

The degeneracy conditions at the particle level, similar to those at the bulk level, are given by:
\begin{equation}\label{deg13}
\bs L_i(\bs z_i)\frac{\partial s_i}{\partial \bs z_i} = \bs 0,
\end{equation}
and
\begin{equation}\label{deg23}
\bs M_i(\bs z_i)\frac{\partial e_i}{\partial \bs z_i} = \bs 0.
\end{equation}

The total energy and entropy of the fluid are obtained by simple addition of the energy and entropy densities over the whole system,
$$
E = \sum_i^{\tt n_v} e_i,
$$
and 
$$
S=\sum_i ^{\tt n_v} s_i.
$$
These conditions ensure that the system evolves in a physically consistent manner, respecting the underlying thermodynamic constraints. 
\begin{figure}[h]
    \centering
    \includegraphics[width=\textwidth]{
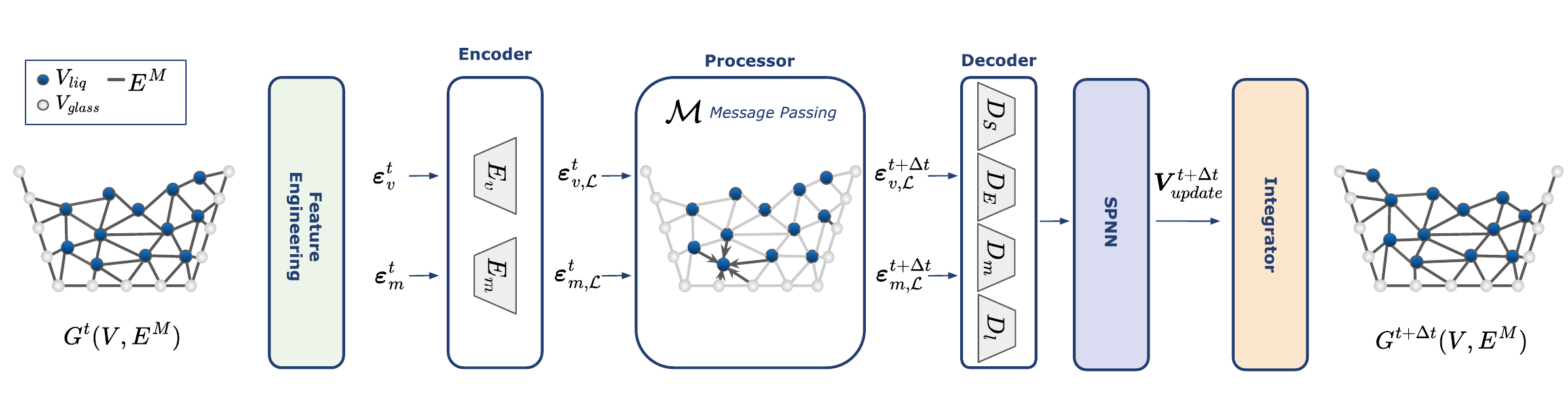}
    \caption{Sketch of the local TIGNN architecture employed for the fluid example.}
    \label{fig:mgn_scheme2}
\end{figure}

\subsubsection{Model architecture}

The architecture of the networks used in this example is very similar to the previous ones, except for the inclusion of the metriplectic bias. In any case, the most important characteristics are reviewed here, see Fig. \ref{fig:mgn_scheme2}.
\begin{itemize}
    \item \textbf{Encoder}: Similar to the one previously seen.
    Two different 2-layer MLPs with 100 hidden units and LeakyReLU activation function are used as encoders. The node encoder $E_v$ for the nodes attributes $\bs \varepsilon^{t}_{v,i}$, and the edge encoder $E_m$ for the edge attributes $\bs \varepsilon^{t}_{m, ij}$.

    \item \textbf{Processor}: The message passing consists on $M=5$ message passing blocks of 2-layer MLPs with 100 hidden units and LeakyReLU activation.
    
    \item \textbf{Decoders}: This section module includes the more significant differences from the previous architecture. Here, we utilize four decoders, each consisting of 2-layer MLPs with 100 hidden units. Two decoders are responsible for predicting the GENERIC energy gradient $\frac{\partial e_i}{\partial \bs z_i}$, and the entropy gradient, $\frac{\partial s_i}{\partial \bs z_i}$, for each particle based on the output from the node block. The remaining two decoders predict the flattened operators $\bs l$ and $\bs m$  for each edge, where the input features include both nodal and edge characteristics.

    \item \textbf{SPNN}: This block is designed to enforce the metriplectic structure of dissipative Hamiltonian systems \cite{hernandez2023port}. It ensures that the conditions of positive semi-definiteness and skew-symmetry are satisfied by imposing the following constraints: 
    \begin{equation}\label{conf_lm}
    \bs M_{ij}=\bs m_{ij}\bs m_{ij}^{\top},   \qquad   \bs L_{ij}= \bs l_{ij}- \bs l_{ij}^{\top}.
    \end{equation}

    \item \textbf{Loss function}: We build the neural network using two loss terms. The first is a data-loss term, which measures how accurately the network predicts the time derivative of the state vector compared to the ground truth, integrated using GENERIC:
    \begin{equation*} 
    \mathcal{L}^{\text{data}} = \left\Vert \dot{\bs{z}}^{\text{GT}} - \dot{\bs{z}}^{\text{net}} \right\Vert^2_2,
    \end{equation*}
    where \(\dot{\bs{z}}^{\text{GT}}\) is the ground truth solution, \(\dot{\bs{z}}^{\text{net}}\) is the network prediction, and \(\Vert \cdot \Vert_2\) represents the L2-norm. This formulation regularizes the global loss by focusing on the time derivative instead of the state vector.
    
    The second loss term softly enforces the fulfillment of the degeneracy equations:
    \begin{equation*}\label{loss_deg}
    \mathcal{L}^{\text{deg}} = \left\Vert \bs{L} \dpar{S}{\bs{z}} \right\Vert^2_2 + \left\Vert \bs{M} \dpar{E}{\bs{z}} \right\Vert^2_2.
    \end{equation*}
    
    The total loss is computed as the sum of these two terms, with a hyperparameter \(\lambda\) to balance their contributions:
    \begin{equation*}\label{loss_data}
    \mathcal{L} = \mathcal{L}^{\text{deg}} + \lambda \mathcal{L}^{\text{data}}.
    \end{equation*}

\item \textbf{Integrator:}  
In the final step, we apply the metriplectic structure as a soft constraint to design a thermodynamically consistent integrator, which acts as the thermodynamic inductive bias of our method. This integration step is performed using a forward Euler scheme with a time step $\Delta t$, incorporating the nodal GENERIC formalism,
\begin{equation}\label{intgration}
\bs z_i (t+\Delta t) = \bs z_i (t) + \dot{ \bs z_i}(t) \Delta t  .
\end{equation}
\end{itemize}

\subsubsection{Experiments}

In this final experiment, we take a network previously trained using 35 simulations of different liquid volumes sloshing around inside a straight-walled cylindrical vessel. The training took just over two days.

When exposed to a new container geometry, the previous model fails after only three time steps, making it unable to perform inference on a container with a slight slope in its walls.

To overcome the lack of generality of the just trained model, we now refine this network from the pre-trained weights, using a new dataset consisting of only 6 additional simulations for training and validation with previously unseen container geometries. Compared to the initial training, which lasted two days, this new training took only 100 minutes.

The state variables include position $\bs q_{j}$, velocity $\bs v_{j}$, and internal energy $ \bs e_{j}$, and the data has been synthetically generated using Smoothed Particle Hydrodynamics (SPH).

For this problem, the characteristic parameters include a wave speed of \( c = 10 \) and a density of \( \rho = 983.2 \), while the shear viscosity is set to \( 1.3 \times 10^{-3} \). The dimensions of the glass vary from a height between \( H = 0.04 \) and \( H = 0.08 \) and a radius between \( R = 0.015 \) and  \( R = 0.070 \). The liquid level in the glass ranges from \( H_{\text{liq}} = 0.015 \) to \( H_{\text{liq}} = 0.030 \), with a varying number of particles, averaging around 6000. The glass is subjected to an impulsive force of \( f_{\text{ini}} = 0.25 \), and the connectivity radius is defined as \( r_c = 0.007 \).

The selected hyperparameters for training involve a batch size of 1 doe to the huge number of nodes and edges, and the learning rate begins at \( lr = 8 \times 10^{-4} \), gradually decreasing at epochs 20, 30, and 40, over a total of \( N_{\text{epoch}} = 50 \) epochs. Furthermore, the variance of the training noise is set to \( \sigma_{\text{noise}} = 8 \times 10^{-4} \).

\begin{figure}[h]
    \centering
    \includegraphics[width=\textwidth]{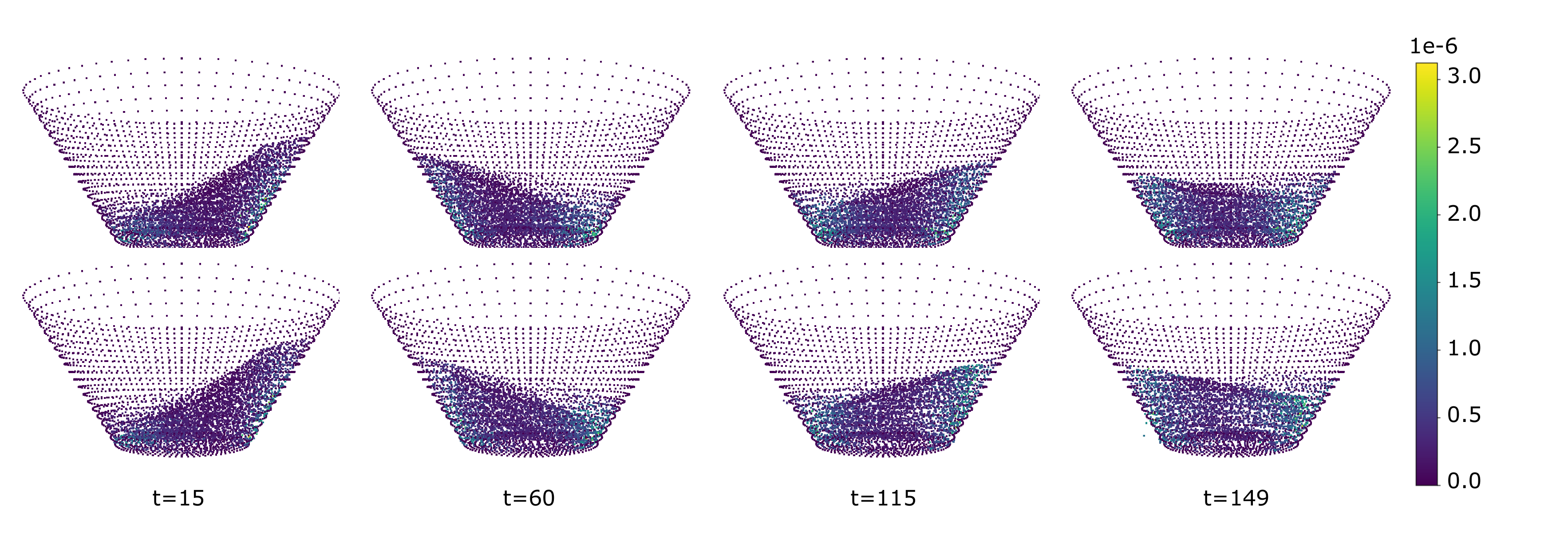}
    \caption{Results for a simulation of  water sloshing in a container with trocho-conical walls. Four snapshots of the sloshing sequence were selected for comparison. The first row corresponds to the fluid reconstruction, while the second row shows the ground truth of the simulation. The colour of the particles indicates the energy level of each node.}
    \label{fig:glass_resutls}
\end{figure}

Figure \ref{fig:glass_resutls} shows a rollout of a simulation with a previously unseen geometry, covering 149 time steps, the same duration used during training. The visualization highlights the network's ability to accurately capture the dissipative behavior of the fluid, progressively bringing it to rest. Additionally, the RMSE and RRMSE metrics for this simulation are provided in Table \ref{tab:rmse_rrmse2}.

\begin{table}[h]
    \centering
    \begin{tabular}{|c|c|c|c|}
        \hline
        \textbf{Variable} & Position& Velocity& Energy\\
        \hline
        RMSE& 3.22 \(\times 10^{-3}\)& 2.34 \(\times 10^{-2}\)& 2.17 \(\times 10^{-7}\)\\
        \hline
        {RRMSE (\%)}& 3.15& 13.6& 0.491\\\hline
    \end{tabular}
    \vspace{0.5cm} 
    \caption{Zero-shot performance RMSE and RRMSE values for position ($\bs q$), velocity ($\bs v$) and energy ($E$) for the simulation test.}
    \label{tab:rmse_rrmse2}
\end{table}

This example demonstrates how a training based on the results of 35 high-fidelity simulations and costing more than two days, being unable to adapt to a major change in container geometry (change from vertical to inclined walls) is generalised with a small amount of transfer learning training using only 6 new simulations on containers of different geometries. This ensures a much higher degree of generality, resulting in a simulator capable of providing results with a very high degree of accuracy for tanks of arbitrary size and geometry, as well as arbitrary fluid volumes.

\section{Conclusions}\label{conclusions}

Foundational models have demonstrated a tremendous capacity to revolutionise the world of artificial intelligence. However, their impact is even more limited in the world of computational simulation of physical phenomena. In this paper we have explored, on the one hand, the ability of two graph network architectures to deal in a zero-shot strategy with out-of-distribution problems. On the other hand, we have developed and analysed different strategies for model refinement in the face of new demands. First, modifications due to the requirement of new output variables in already known phenomena. This has been achieved thanks to a transfer learning strategy. Finally, a one-shot strategy has been developed to adapt the models to very severe changes in the geometry.

In all cases studied, modifications to models trained on reasonably complex datasets (consisting of high-fidelity simulation results) have proven to have a very limited computational cost.

It should be noted that in all cases the initial training has been performed on datasets that, although large, in no way correspond to the standards currently used for natural language processing applications, for example. This allows us to ensure that the use of even larger datasets will allow us to obtain even more accurate results, although this is beyond the scope of the means available to the authors. 

In sum, this paper is intended as a first exploration of what foundational models can contribute to the world of computational simulation of physical phenomena. We believe that there is immense room for improvement and that our community has a vast territory to explore. We believe, however, that the proposed architectures, without prejudice to the study of others or the improvement of those already analysed, offer a capacity for generalisation to previously unseen situations and particularly out of distribution that allow us to be optimistic about the future of this type of technique.

\section*{Acknowledgements}

This work was supported by the Spanish Ministry of Science and Innovation, AEI/10.13039/501100011033, through Grants number TED2021-130105B-I00  and PID2023-147373OB-I00, and by the Ministry for Digital Transformation and the Civil Service, through the ENIA 2022 Chairs for the creation of university-industry chairs in AI, through Grant TSI-100930-2023-1.

This research is also part of the DesCartes programme and is supported by the National
Research Foundation, Prime Minister Office, Singapore under its Campus for Research
Excellence and Technological Enterprise (CREATE) programme.

This material is also based upon work supported in part by the Army Research Laboratory and the Army Research Office under contract/grant number W911NF2210271.

The authors also acknowledge the support of ESI Group through the chairs at the
University of Zaragoza and at ENSAM Institute of Technology.


\end{document}